\documentclass{ws-p8-50x6-00}
\newcommand{\open}{< \kern -0.3em \scriptscriptstyle{)}}
\def\NPA{\em Nucl. Phys. A}

\begin{document}

\title{Interference Fragmentation Functions and Spin Asymmetries}

\author{M. Radici}

\address{Dipartimento di Fisica Nucleare e Teorica, Universit\`a di Pavia, and \\
Istituto Nazionale di Fisica Nucleare, Sezione di Pavia, I-27100 Pavia, Italy}


\maketitle

\abstracts{A new class of fragmentation functions, arising from the interference of 
different hadron production channels, is analyzed in detail. Their symmetry 
properties with respect to naive time-reversal transformations allow for 
the exploration of final-state interactions occurring during and after the 
hadronization. Their symmetry properties with respect to chiral transformations 
allow building spin asymmetries where the quark transversity distribution can be
factorized out at leading twist. Explicit calculations will be shown for the 
interference fragmentation functions arising from final-state interactions of two 
pions detected in the same current jet for the case of semi-inclusive Deep-Inelastic 
Scattering (DIS).}

Because a rigorous explanation of confinement is still far from being achieved,
information on the nonperturbative nature of quarks and gluons inside hadrons can be
extracted from distribution (DF) and fragmentation (FF) functions in hard scattering
processes. There are three independent DF that completely determine the quark status
inside hadrons at leading twist with respect to its longitudinal momentum and spin:
the momentum distribution $f_1$, the helicity distribution $g_1$ and the
transversity distribution $h_1$. In contrast to the first two ones, $h_1$ is related
to soft processes that flip chirality; as such, $h_1$ is, e.g., unaccessible in 
inclusive DIS. However, the new generation of machines
(HERMES, COMPASS, eRHIC) allows for a better resolution in the final state and
precise semi-inclusive measurements are becoming feasible. In this context, naive 
time-reversal odd (for brevity, ``T-odd'') FF naturally arise because no constraints
from time-reversal invariance can be imposed due to the existence of Final State 
Interactions (FSI) with or inside the residual jet~\cite{ruju}. 
The interference of different hadron production channels produces a new set of T-odd 
FF which are also chiral odd, and, therefore, they represent the natural partner to 
isolate $h_1$ already at leading twist, whereas other tools like Drell-Yan processes
seem less favoured~\cite{boer}.

From the theoretical point of view, it is convenient to select DIS processes 
where two unpolarized leading hadrons with momenta $P_1$ and $P_2$ are detected in 
the same jet, that acts as a spectator~\cite{jaffe,noi1}, and generalize the 
Collins-Soper 
light-cone formalism~\cite{collsop} to this case. At leading twist, four FF appear: 
$D_1, G_1^\perp, H_1^\perp, H_1^{\open}$~\cite{noi1}. In the frame where the
transverse component of the 3-momentum transfer vanishes, ${\bf q}_\perp=0$, they 
depend on how much of the fragmentig quark momentum $k$ is carried by the hadron 
pair ($z=(P_1+P_2)^-/k^-\equiv P_h^-/k^-=z_1+z_2$), on the way this momentum is 
shared inside the pair ($\xi=z_1/z$), and on the ``geometry'' of the pair, namely on 
its transverse relative momentum (${\bf R}^2_\perp$, with $R=(P_1-P_2)/2$), on the 
relative position of the pair total momentum $P_h$ with respect to the jet axis 
${\bf \hat k} = {\bf k} / |{\bf k}|$ (i.e. ${\bf k}^2_\perp$), and on the relative 
position between the pair plane, formed by $P_1$ and $P_2$, and the plane containing 
$P_h$ and the jet axis (i.e. ${\bf k}_\perp \cdot {\bf R}_\perp$). Each FF is also
related to a specific spin state of the fragmenting quark: $H_1^\perp$ is the
analogue of the Collins function for the one hadron semi-inclusive DIS~\cite{coll}; 
on the contrary, $H_1^{\open}$ represents a genuine new effect relating 
the transverse polarization of the fragmenting quark to the transverse relative 
dynamics of the detected pair, namely ${\bf R}_\perp$. $G_1^\perp, H_1^\perp, 
H_1^{\open}$ are ``T-odd'' and are nonvanishing only in the presence of residual 
FSI, at least between the two hadrons. $G_1^\perp$ is chiral even, while both 
$H_1^\perp, H_1^{\open}$ are chiral odd and can be identified as the chiral partner 
needed to access the transversity $h_1$~\cite{noi1}. 

\begin{figure}[t]
\epsfxsize=12cm 
\epsfbox{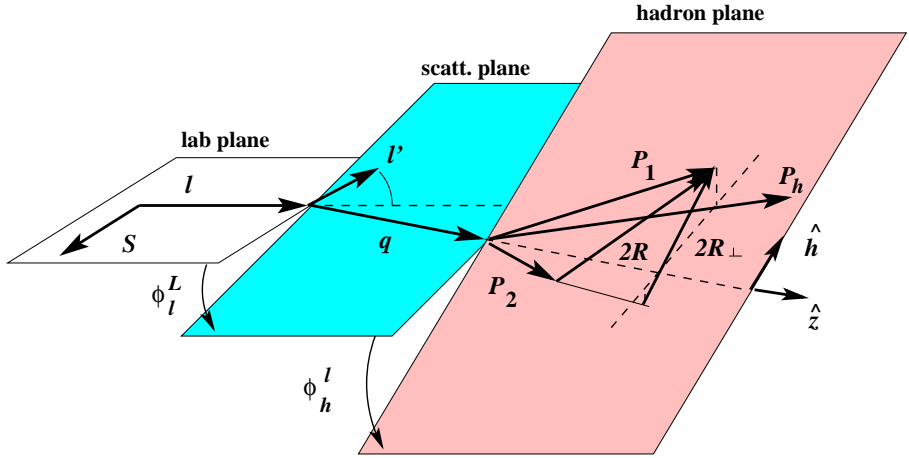} 
\caption{Kinematics for semi-inclusive DIS where two leading hadrons are detected
(see text). \label{fig1}}
\end{figure}

The cross section at leading twist for the process $eN\rightarrow e'h_1h_2X$ has 
been worked out in detail in Ref.~\cite{noi1}. Here, we will reconsider the case for 
an unpolarized beam and a transversely polarized target. The lab frame can be 
defined by the plane containing the beam 3-momentum 
${\bf l}$ and the target polarization ${\bf S}$ (see Fig.~\ref{fig1}). The 
scattering plane, which contains the scattered lepton 3-momentum ${\bf l}'$ and 
${\bf l}$, is rotated by the azimuthal angle $\phi_l^L$. Finally, the socalled 
hadronic plane, which contains ${\bf q}={\bf l}-{\bf l}'$ (conventionally parallel 
to the ${\bf \hat z}$ axis) and ${\bf P}_h$, is rotated by 
$\phi_h^L = \phi_h^l+\phi_l^L$. A further plane, which is just sketched in 
Fig.~\ref{fig1} for sake of simplicity, contains ${\bf P}_1, {\bf P}_2, {\bf R}$ and 
its transverse component ${\bf R}_\perp$; it is rotated by 
$\phi_R^L=\phi_R^l + \phi_l^L$ with respect to the lab. 
The nine-fold differential cross section depends on the energy fraction taken by the
scattered lepton ($y=q^0/|{\bf l}|$), on $\phi_l^L$, on the quark light-cone 
momentum fraction $x=p^+/P^+$ of the target momentum $P$, on $z, \xi, {\bf R}_\perp$ 
and  ${\bf P}_{h\perp}$. Since $R_\perp^2 = \xi (\xi-1) P_h^2 -(1-\xi) P_1^2 -\xi 
P_2^2$~\cite{noi1}, the cross section
can be more conveniently considered differential with respect to the pair invariant 
mass $P_h^2=M_h^2$ and $\phi_R^L$. By integrating over the ``internal'' dynamics
(i.e. on $\xi, {\bf k}_\perp, {\bf P}_{h\perp}$) and by properly folding the cross
section over the experimental set of beam and hadron-pair azimuthal positions
$\phi_l^L$ and $\phi_R^L$, it is possible to come to the factorized expression
\begin{eqnarray}
\langle d\sigma_{OT} \rangle   
&\equiv &\int d\phi_l^L \, d\phi_R^L d{\bf P}_{h\perp} \, d\xi \, 
         \sin (\phi_R^L - 2 \phi_l^L ) 
    \frac{   d\sigma_{OT}  }
         {dy \, d\phi_l^L \, dx \, dz \, d\xi \, dM_h^2 \, d\phi_R^L \, 
	   d{\bf P}_{h\perp}  }   \nonumber \\
&=  &\frac{\alpha_{em}^2 s}{4\pi^2 Q^4} \, \frac{(1-y) |\vec S_\perp |}{M_1+M_2} 
     \, \sum_a e_a^2 x \, h_1^a (x) \, H_1^{\open \, a}
     (z,M_h^2)
\label{eq:finalx}
\end{eqnarray}
where $\alpha_{em}$ is the electromagnetic fine structure constant, $s=Q^2/xy$ is
the total energy in the center-of-mass frame, $M_1,M_2$ are the masses of the two
observed hadrons, and a sum over the flavor $a$ of each quark with charge $e_a$ is
performed (see also Ref.~\cite{physrep}). It is worth noting that such a factorized 
form can be reached requiring one less variable than in the case of one hadron 
emission with the Collins functions, where also knowledge of $|{\bf P}_{h\perp}|$ is 
needed~\cite{boer}.

\begin{figure}[t]
\epsfxsize=7cm 
\hspace{2cm} \epsfbox{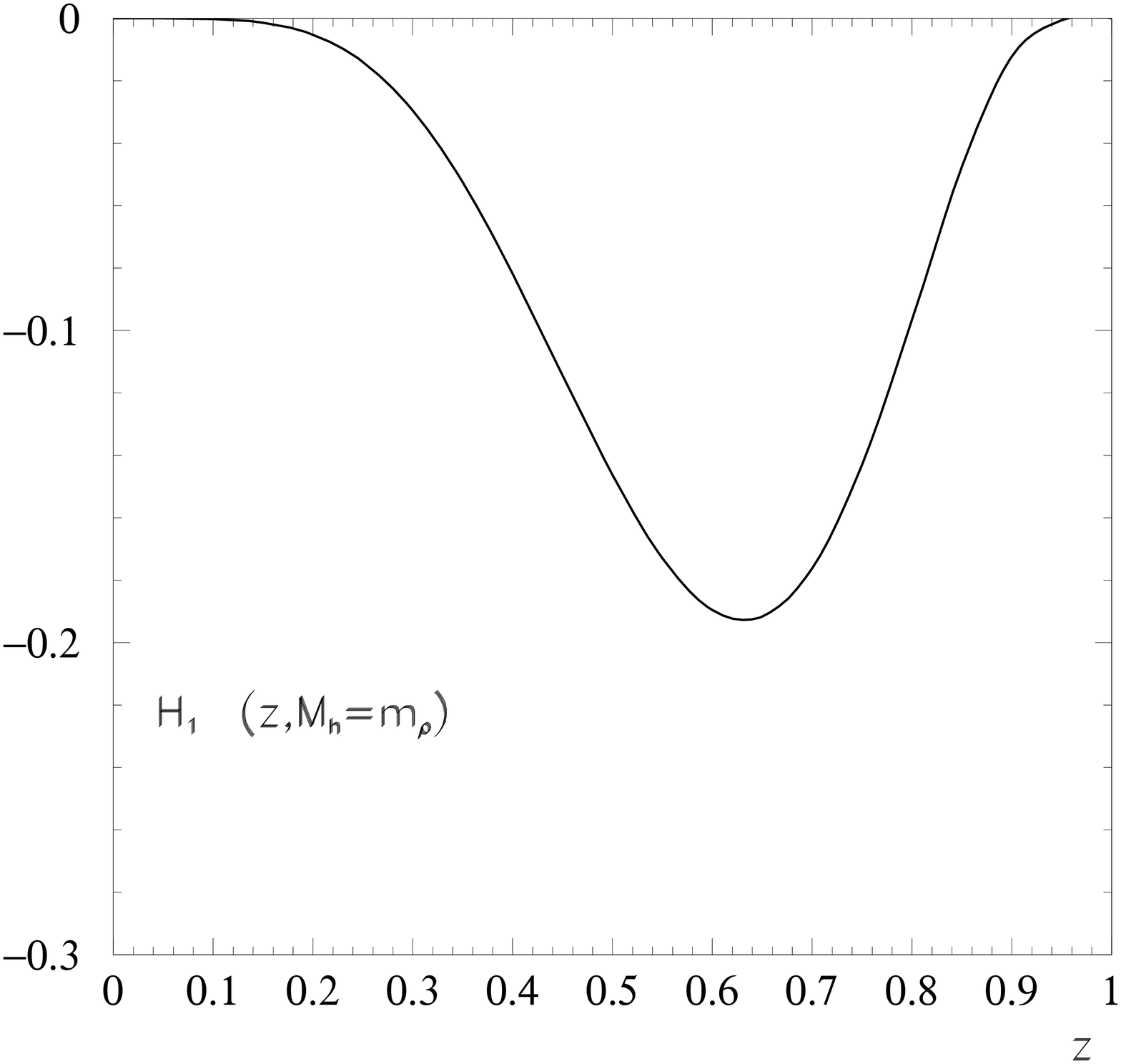} 
\caption{The fragmentation function $\alpha_{em}^2 / [8\pi^2 m_\pi] \times 
H_1^{\open} (z,M_h^2=m_\rho^2)$ of Eq.~(\ref{eq:finalx}) for the 
$u\rightarrow \pi^+ \pi^-$ case. \label{fig3}}
\end{figure}

Quantitative predictions for the integrated $H_1^{\open \, a}$ in 
Eq.~(\ref{eq:finalx}) can be produced by specializing the spectator model of 
Ref.~\cite{jakob} to the case of the emission of a hadron pair. For the hadron pair 
being a proton and a pion, results have been published in Ref.~\cite{noi2}, where 
FSI arise from the interference between the direct production and the Roper decay. 
Here, results are shown for the case of two pions with invariant mass in the range 
$[m_\rho-\Gamma_\rho, m_\rho+\Gamma_\rho]$, with $m_\rho=768$ MeV and 
$\Gamma_\rho\sim 250$ MeV. The spectator state has the quantum numbers of an 
on-shell quark with constituent mass $m_q=340$ MeV. Interference ``T-odd'' FF arise 
from the interference between the direct production  and the $\rho$ decay. In 
Fig.~\ref{fig3}, $\alpha_{em}^2 / [8\pi^2 m_\pi] \times H_1^{\open} 
(z,M_h^2=m_\rho^2)$ of Eq.~(\ref{eq:finalx}) is shown for the case 
$u\rightarrow \pi^+ \pi^-$. It shows that a nonvanishing integrated interference FF 
survives allowing for the extraction of $h_1$ at leading twist, as suggested in 
Eq.~(\ref{eq:finalx}).

\section*{Acknowledgments}
This work has been done in collaboration with A. Bianconi (Univ. Brescia), D. Boer
(RIKEN-BNL) and R. Jakob (Univ. Wuppertal).

\end{document}